\documentclass[conference]{IEEEtran}
\usepackage{graphicx}
\usepackage{amsmath}
\ifCLASSINFOpdf
\else
\fi
\begin{document}
%
\title{Scalable Self-Adaptive Synchronous Triggering \\ System in Superconducting Quantum Computing}

\author{\IEEEauthorblockN{Li-Hua Sun, Fu-Tian Liang, Jin Lin, Cheng Guo, Yu Xu, Sheng-Kai Liao and Cheng-Zhi Peng}
\IEEEauthorblockA{Hefei National Laboratory for Physical Sciences at the Microscale and Department of Modern Physics,\\
University of Science and Technology of China, Hefei 230026, China\\
Chinese Academy of Sciences (CAS) Center 
for Excellence and Synergetic Innovation Center in\\
 Quantum Information and Quantum Physics, 
University of Science and Technology of China, Shanghai 201315, China\\
Email: sunlihua@mail.ustc.edu.cn
}}


%


\maketitle

\begin{abstract}
Superconducting quantum computers (SQC) can solve some specific problems which are deeply believed to be intractable for classical computers. The control and measurement of qubits can't go on without the synchronous operation of digital-to-analog converters (DAC) array and the controlled sampling of analog-to-digital converters (ADC). 
In this paper, a scalable self-adaptive synchronous triggering system is proposed to ensure the synchronized operation of multiple qubits. 
The skew of the control signal between different qubits is less than 25~ps.
After upgrading the clock design, the 250~MHz single-tone phase noise of DAC has been increased about 15~dB. 
The phase noise of the 6.25~GHz qubit control signal has an improvement of about 6~dB.

\end{abstract}


%
\IEEEpeerreviewmaketitle

\section{Introduction}


Superconducting quantum computer (SQC) is composed of a number of superconducting qubits which can be manipulated by microwave signals \cite{Nielsen2000}\cite{Martinis2004}.
Currently, the control and measurement of superconducting qubits are performed in a dilution refrigerator, in which a coordinate sequence of microwave pulses is interact with the qubits through coaxial cables \cite{Sank2014}. 
With regard to the preparation and recording of microwave pulses, the current practice is to use a radio-frequency signal mixing up or down with a low-frequency microwave signal which can be generated and sampled by 1G/s DACs and ADCs for general \cite{Chen2012}. 

In distributed system, synchronization is often indispensable and difficult \cite{Tanenbaum2006}. 
The electronic operation system of SQC is such a kind of special distributed data converters array due to both the distributed control and measurment framework and the large quantity of qubits, and thus it also requires a good synchronization method so as to achive the high-precision synchronization of qubits. 
In addition, because the state of qubit is sensitive to the frequency and phase of microwave, the microwave modulation signal must be stable and controllable, that is to say, the phase noise should be as small as possible. 

In this paper, in order to achieve a high-stability and high-precision qubit synchronizatin operation for SQC, a scalable and self-adaptive synchronous triggering system is proposed. 
The main contributions of the article are as follows: 
(1) Upgrade the clock system to improve the performance of qubit control signals. 
(2) Propose a star-like triggering system to realize the synchronous manipulation of multiple qubits.
(3) Design self-adaptive program to solve the metastable problem caused by the master-salve trigger mechanism.



\section{Control and measurement system of sqc}
\begin{figure}[!h]
\centering
\includegraphics[width=3.5in]{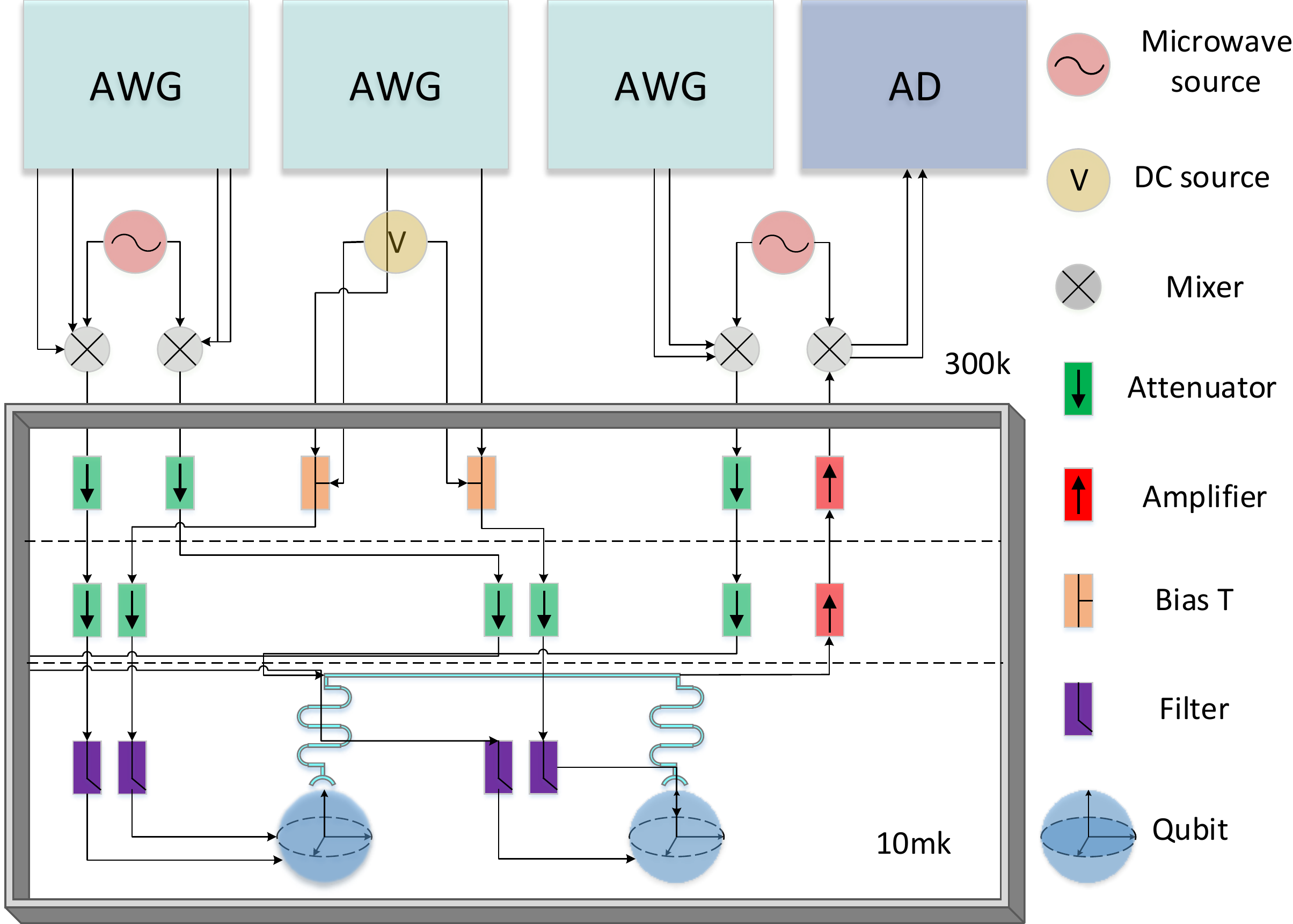}A
\caption{The experiment equipment of SQC. }
\label{SQC}
\end{figure}

The schematic of SQC control and measurement system is shown in Fig. \ref{SQC} which exhibits two qubits as the example. 
The qubits work at an ambient temperature below 10mk that is provided by a dilution refrigerator \cite{Sank2014}. 
In the process of control, the modulating microwave signal from room temperature is transmitted to qubit after multistage refrigeration, attenuation and filtering. 
These modulating signals can be obtained by mixing up the DC-250~MHz signals generated by the DAC-based arbitrary waveform generator (AWG) with the 4-8~GHz RF signals provided by the microwave sources. 
In the process of measurement, a modulating microwave signal is first fed into the resonant cavity coupled with the qubits. 
Due to different states of qubits corresponding to the different amplitudes and phases of the microwave signal on the resonator, 
the state information of qubits can be obtained from the amplitudes and phases which can be sampled by ADC after feedbacking from resonator. 

As shown in Fig. \ref{SQC}, controlling a qubit requires two modulating microwave signals: 
one is obtained by mixing up the two DC-250~MHz signals produced by self-developed AWG and a 4-8~GHz RF signal generated from the microwave source via an IQ mixer; 
another is synthesized by a Bias T using a sequence of pulses generated from AWG and an ultra-low ripple DC signal provided by high-precision DC source. 
With respect to the measurement of qubits, several qubits can share a reading route with a ratio of about 4:1 \cite{Zheng2017}. 

Taking a 4-channel AWG and 2-channel ADC as an example, the number of qubits and the required AWG and ADC resources are shown in table \ref{table1}. 

\begin{table}[!h]
\renewcommand{\arraystretch}{1.5}
\caption{The Number of Qubits and the Required AWGs and ADCs}
\label{table1}
\centering
\begin{tabular}{|c|c|c|c|}
\hline
Object & Qubits & AWGs & ADCs\\
\hline
Required Number & N & 4N & N/4\\
\hline
\end{tabular}
\end{table}

With hundreds of qubits in the future, the number of AWGs and ADCs to be synchronized will increase linearly, which requires a better scalability of the synchronization system. 

\section{Design of Synchronizing Trigger System}

\subsection{The High-performance Clock System}

\begin{figure}[h]
\centering
\includegraphics[width=3.5in]{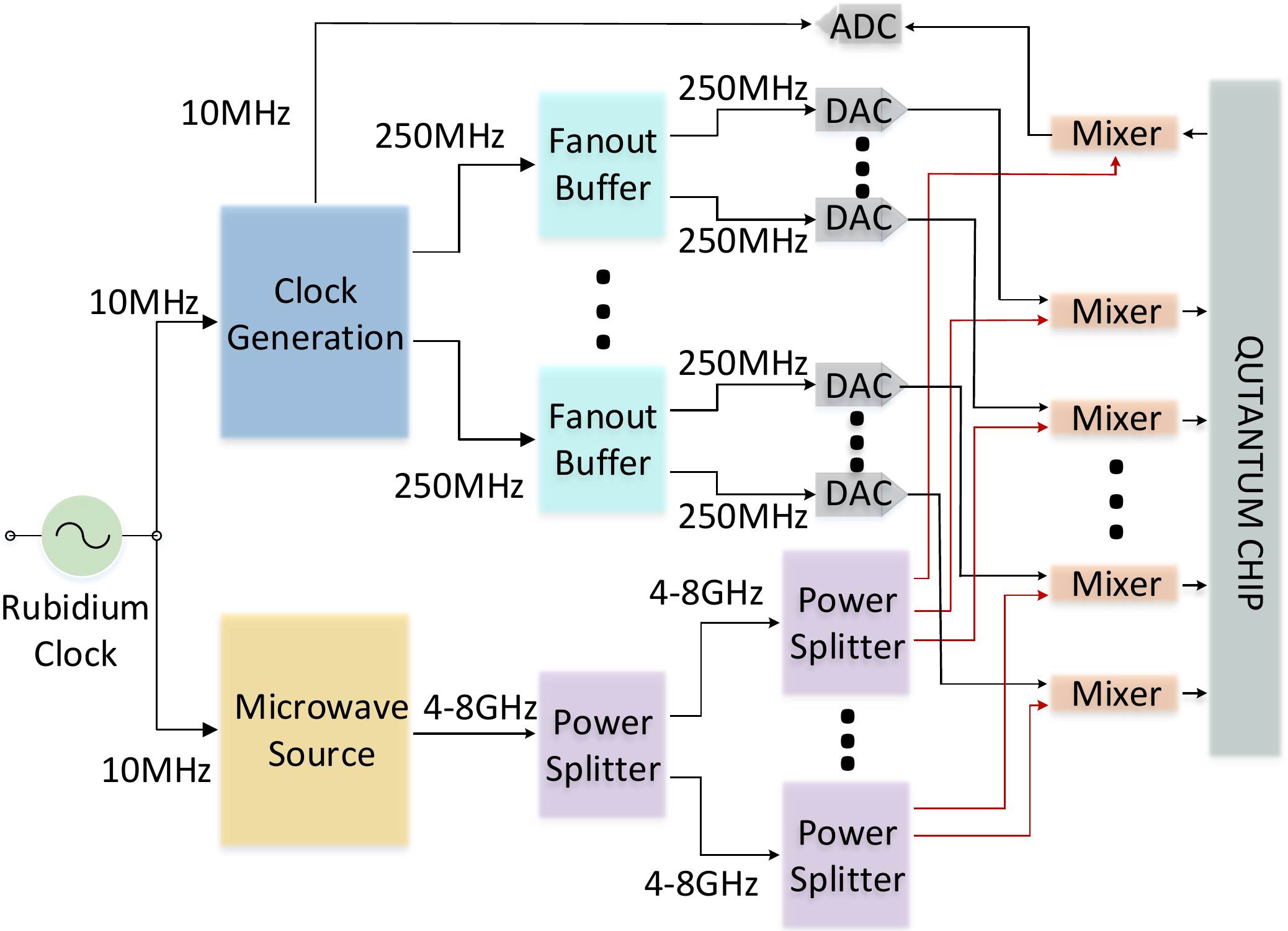}
\caption{The clock structure of the synchronous triggering system. }
\label{clock}
\end{figure}

In our original clock scenario, the homologous 10~MHz clock from a rubidium clock source generates a 2~GHz clock via the AD9516 clock IC in each AWG. 
And the 2~GHz clock is divided into several 250~MHz clocks to work as the synchronization clock. 
This scheme results in a skew of 0.5~ns between different AWGs and a limited number of synchronizable AWGs

In addition, according to Ball's test \cite{Ball2016}, when the phase noise of the main clock that control the qubits is decreased by 40 dB, the fidelity of the qubits can be increased by 3 orders of magnitude. 
Therefore, the clock performance should be as good as possible under the premise of balanced parameters.  

In order to solve the problem of the original design and imporve the clock performance, we propose a high-performance tree-based homologous clock system. 
In the clock system, a 10~MHz Rubidium clock is deployed as the root of the clock tree for the entire SQC system. 
The rubidium clock has multiple coherent 10MHz clock signals, which are mainly used in three aspects (Fig. \ref{clock}) : 
the first one is locked to 250~MHz by a high-performance clock gengertor HMC7044 and fanned out to to each AWG as the synchronous reference clock through the fan-out buffers HMC7043 with ultra-low phase noise; Since HMC7044 and HMC7043 each has 28 single-ended outputs, the clock tree with one level can support up to 784 AWGs' synchronization.
the second one is used as the reference clock of microwave source which is applied to generate the 4-8~GHz RF signal. 
Similarly, the RF signal is divided into multiple channels as the local inputs of IQ mixers through the power splitter of multi-level tree-like struture. 
The last one is used as the clock source for ADC sampling unit. 
 
This upgraded design ensures that each AWG obtains a synchronized clock with the excellent synchronization skew and a low phase noise. 
The related test results will be presented in the following sections. 

\subsection{The Star-like Trigger Design}
During the operation of qubits, usually after one qubit manipulation is completed, if the qubit is in an excited state, it needs to wait until it decays to the ground state before the next operation can be performed. 
The decay time is proportional to the qubit decoherence time, usually for no more than 200~us. 
This natural decay method greatly reduces the operating efficiency of the qubit. 
Therefore, it is necessary to artificially add a reset operation to initialize the state of qubit to the ground state, and the artificial reset time is less than 1us, which can significantly improve the operating efficiency \cite{Riste2015}\cite{Liu2016}.
 
\begin{figure}[h]
\centering
\includegraphics[width=2.6in]{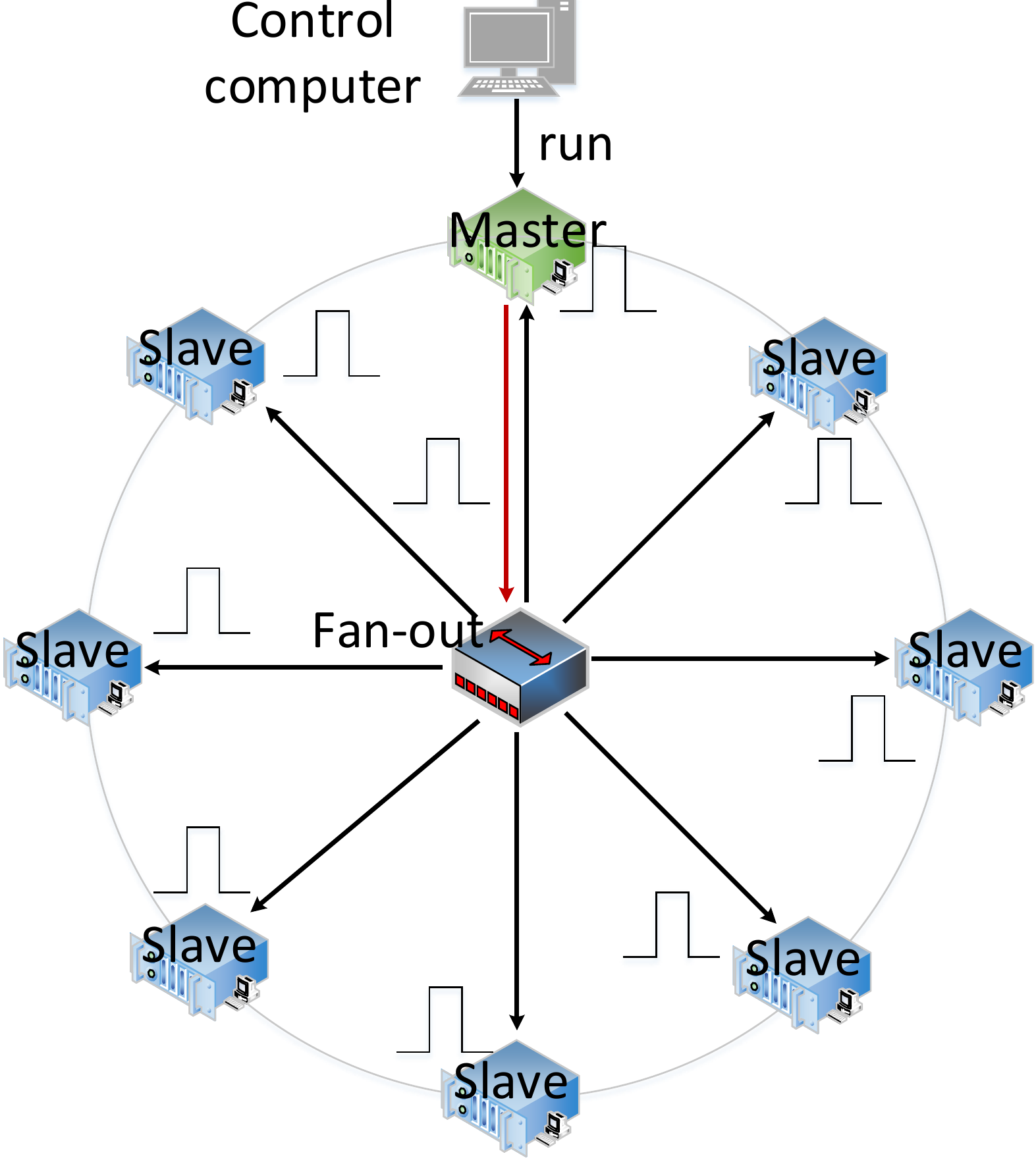}
\caption{The star-like fanning out structure}
\label{star}
\end{figure}

\begin{figure*}[h]
\centering
\includegraphics[width=6.5in]{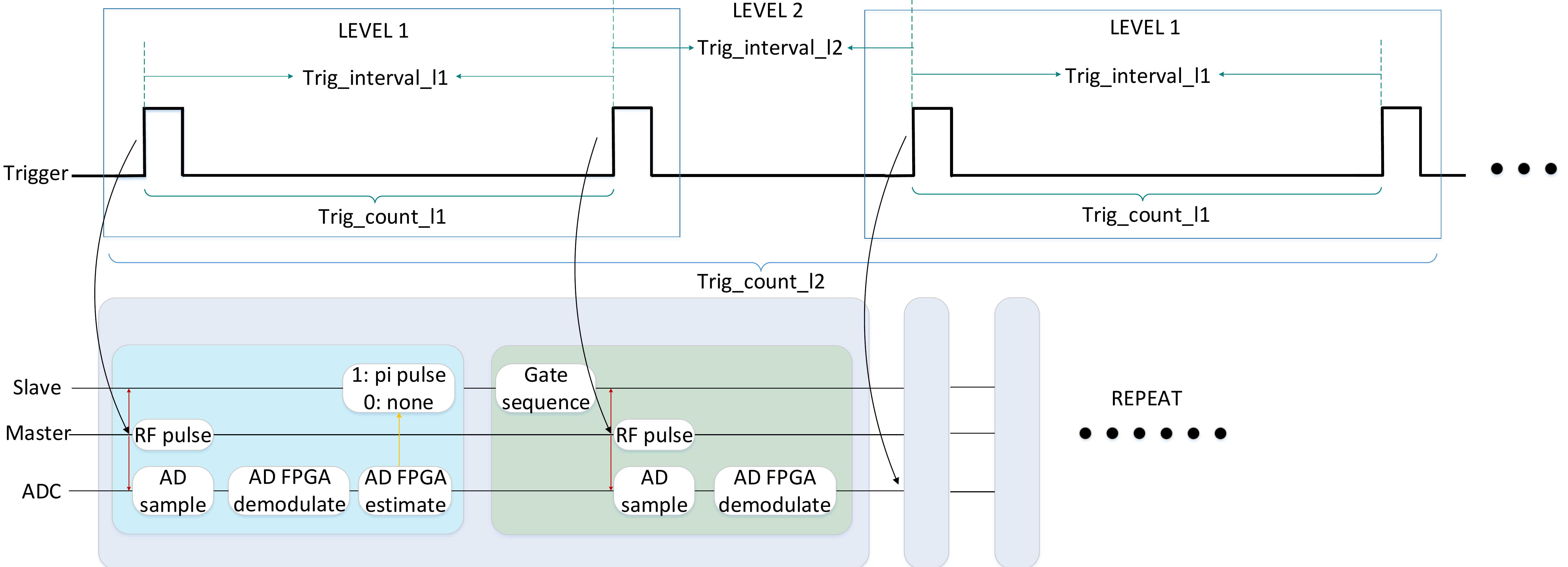}
\caption{The timing diagram of triggering mechanism}
\label{trig}
\end{figure*}

The clock system described in previous subsection ensures that each AWG has a synchronous sampling clock. 
Here we propose an FPGA-based two-level trigger system, which can provide synchronous trigger pulses for each AWG and support the reset operations.
This trigger system is a star-like structure based on the master-slave model.
As shown in Fig. \ref{star}, one of the AWG is set as the master and the others are set as slaves. 
When the master receives a 'run' instruction from control-computer, it outputs a series of trigger pulses, with configurable repetition counts, duration times, output delay and intervals. 

The trigger pulses from master will be fanned out by the star-like unit, which is hardware-compatible with the clock tree-based structure. 
And just like the clock system, the number of fan-out triggers can be extended easily by adding fan-out units. 

Fig. \ref{trig} shows the role of each trigger pulse during the reset operation of a qubit. 
The first trigger in level 1 instructs that the ADC measures the initial state of the qubit and feeds it back to the slave AWG. 
At the second trigger moment in level 1, the slave AWG decides whether to emit a $\pi$ pulse for qubit reset based on the received qubit state. 
This is a complete operation with a qubit reset. 
Level 2 is the repetition of the operation in level 1. 
All the intervals and counts involved in Fig. \ref{trig} can be configured as required. 


\subsection{The Self-adaptive Design}

\begin{figure}[b]
\centering
\includegraphics[width=3.5in]{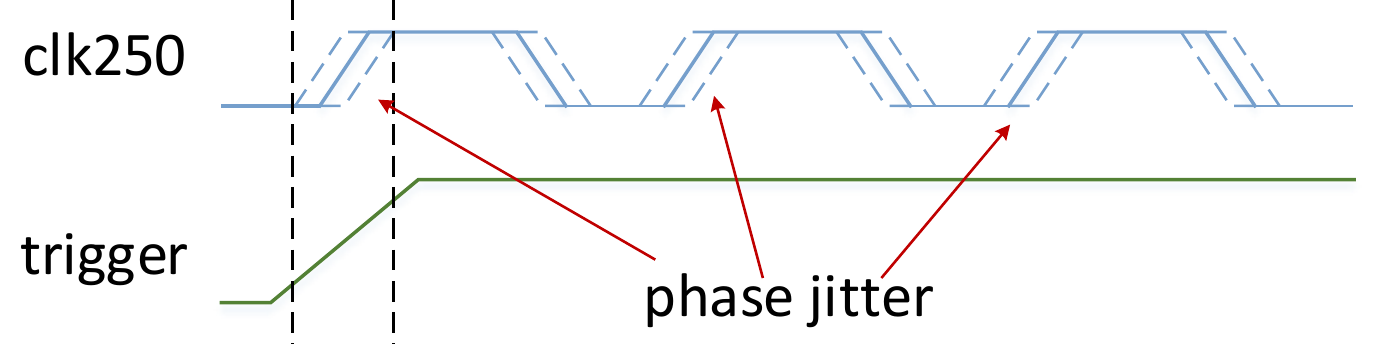}
\caption{The timing diagram between the 250MHz sampling clock and the trigger pulse in FPGA when the metastable phenomenon occurs}
\label{clk250}
\end{figure}

\begin{figure}[]
\centering
\includegraphics[width=3in]{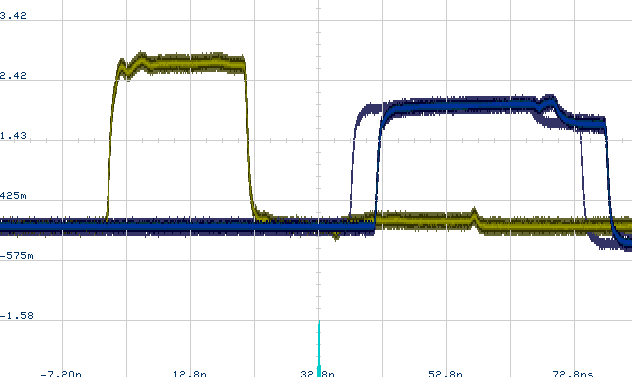}
\caption{The outputs of two AWGs: the yellow one is the output of AWG1 and the trigger source of oscilloscope; the blue one is the output of AWG2. There is a 4~ns jitter between the two AWG outputs}
\label{awg output}
\end{figure}

Under the master-slave trigger mechanism, trigger pulses from the master board are fanned out to each AWG through the fan-out unit. 
When the 250~MHz sampling clock detects the arrival of trigger pulse, the output of AWG is activated. 
However, due to the routing delay of trigger pulse, the 250~MHz clock may sample exactly the rising edge of the trigger, resulting in a metastable phenomenon(Fig. \ref{clk250}). 
When the metastability occurs, the output among multiple AWGs may appear to be out of sync(Fig. \ref{awg output}). 
And once the metastability occurs, it happens frequently. 

The most common solution for metastability is to manually adjust the length of the trigger signal line. 
However, this method is time-consuming and cannot be effectively implemented in large-scale systems. 
Therefore, we propose a self-adaptive method which can automatically detect the meta-stability and calibrate it. 



Our method is based on the IDELAYE and ODELAYE which are a kind of delay resources in FPGA.
In Xilinx Kintex Ultrascale FPGA, the IDELAYE/ODELAYE contains a 512-tap delay line through which the input and output signals of FPGA can be delayed with one tap resloution about 2.5-15~ps. 
What's more, cascading is required when a delay is greater than 1.25~ns \cite{Xilinx2018}\cite{Xilinx2015}. 
In our design, 2 ODELAYE3 and 2 IDELAYE3 are chained together to form an adjustable delay unit that has a maximum delay of about 8~ns. 

\begin{figure}[h]
\centering
\includegraphics[width=3.5in]{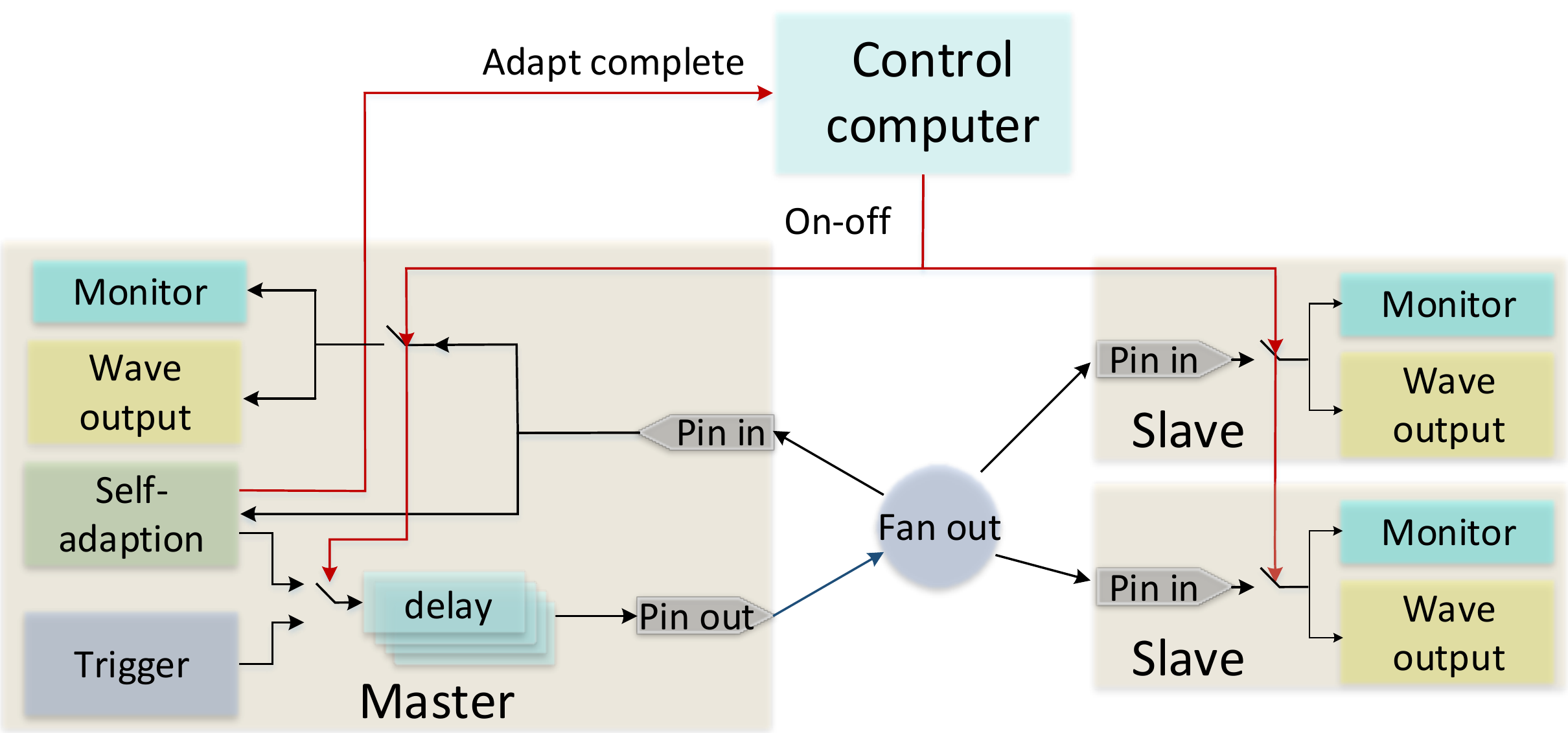}
\caption{A simple reference diagram of signal direction and connection in self-adaptive mechanism}
\label{self_adaption_mechanism}
\end{figure}

\begin{figure}[h]
\centering
\includegraphics[width=3.5in]{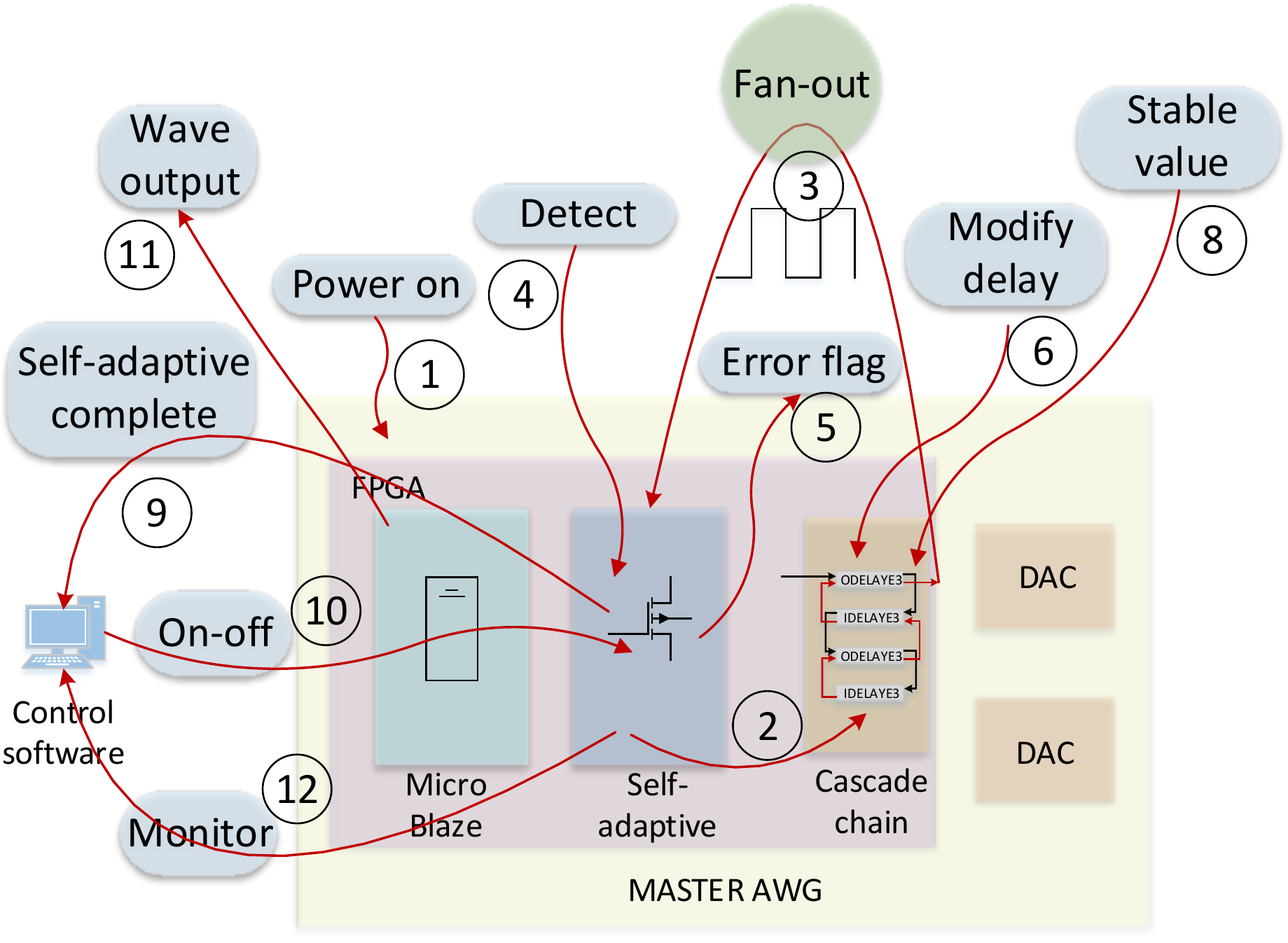}
\caption{A reference diagram of the self-adaptive process(Since step 7 is a repetition of steps 3, 4, 5 and 6, it is not listed here) }
\label{self-adaptive}
\end{figure}
 
Combined with Fig. \ref{self_adaption_mechanism} and Fig. \ref{self-adaptive}, the entire process of self-adaption can be described by the following steps:

 (1) When the master is powered up, the microblaze in FPGA automatically sends an instruction to start the self-adaptive process;

 (2) Master resets the cascade delay chain to the minimum delay;

 (3) Through the cascade chain, the self-adaptive module of master sends out 50000 trigger pulses with an interval of 40 ns. These pulses are returned back to the self-adaptive module of master after being fanned out. Slave AWGs do not respond to these pulses at this time;

 (4) The self-adaptive module of master detects the returned trigger pulses and records the detected intervals and counts;

 (5) The self-adaptive module of master estimates whether the received trigger intervals are equal to the emitted trigger intervals. If not, set the error flag to 1 and record the error counts, otherwise set the error flag to 0;

 (6) Master modifies the cascade delay chain by adding a tap on top of the original delay;

 (7) Master repeats steps 3,4,5 and 6 for 512 times;

 (8) The self-adaptive module of master finds the stable region where metastability does not occur with the help of 512 error flags, and sets the final tap number of the cascade delay chain to the middle value of the stable region;

 (9) Master sends a sign of self-adaptive completion to the control computer;

 (10) The control computer sends on-off instructions to all AWGs, instructing that the switch of the trigger model, the waveform output module and the monitoring module is on and the switch of the self-adaption module is off;

 (11) All AWGs activate the DAC sample to output waveforms based on the triggering pulses;

 (12) The monitoring module detects the received triggers, records the intervals and counts, and estimates whether the received trigger intervals are equal to the configured trigger intervals. If not, the AWG sends a jitter flag to the control computer.

\begin{figure}[h]
\centering
\includegraphics[width=3.2in]{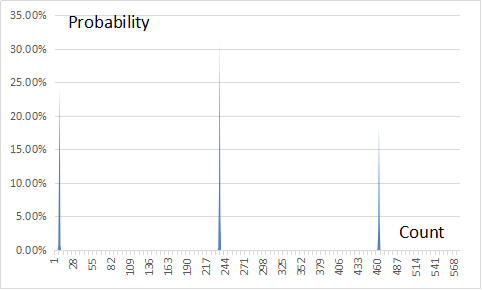}
\caption{The x-coordinate is the number of the control computer reading, corresponding to each modification of the cascade delay chain; the y-coordinate is the probability of metastability. (Because the route delay of control computer reading the number of metastability is relatively large, here we set the trigger intervals to 800~ns and the trigger counts to $10^6$ in the self-adaptive module for testing purpose) }
\label{probability}
\end{figure}

In the process of self-adaption, along with each change of the cascade delay chain, Fig. \ref{probability} records the probability of metastability. 

From the test results of Fig. \ref{probability}, it can be concluded that when the cascade delay chain is continuously modified about 225 times, a clock cycle of 4 ns is traversed, of which about 5 consecutive modifications have a high probability of metastability. 
From this we can conclude that under the triggering mechanism proposed in this paper, the metastability range in a 4~ns clock cycle is about 
$$\dfrac{4000}{225}~*~5~\approx~90~ps$$

In addition, since we use four IDELAYE/ODELAYEs cascade together, it can be calculated that a delay step of an IDELAYE/ODELAYE for one modification is about $$\dfrac{4000}{225~*~4}\approx~4.4~ps$$

By manually adjusting the length of external connection cable for triggers, we build a metastable state and test the self-adaptive function. 
In a total testing of 72 hours, the metastable phenomenon does not appear. 

\section{Test results and analysises}
\begin{figure}[h]
\centering
\includegraphics[width=3in]{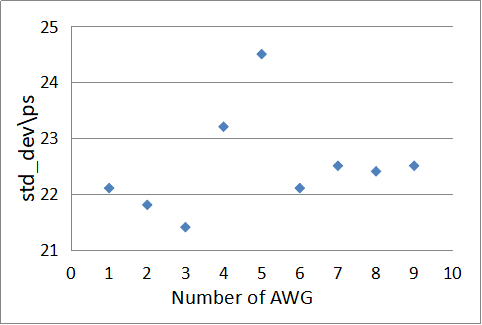}
\caption{The synchronization skew between 10 AWGs}
\label{skew}
\end{figure}

In order to verify the validity of the proposed synchronization triggering method, we test it with 10 AWGs. As shown in Fig. \ref{skew}, the synchronizatin skew among the outputs of 10 AWGs is no more than 25~ps which satisfy the qubit synchronization requirements very well. 

\begin{figure}[h]
\centering
\includegraphics[width=3.3in]{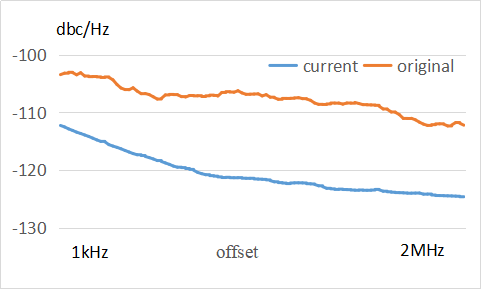}
\caption{Phase Noise Comparisioin of DAC output}
\label{phase noise}
\end{figure}

A comparison test using the original clock design and the upgrated high-performance clock solution is presented here. 
Fig. \ref{phase noise} shows the comparison of the phase nosie of a 250~MHz sine wave output from the same AWG. It can be concluded that the upgrated clock design has a phase noise improvement of about 15~dB in an offset range of 2~kHz to 1~MHz compared to the original clock design. 
Under this test condition, the SFDR of DAC output for both clock schemes are around 50~dB, which means that the clock scheme proposed in this paper improves the phase noise of AWG without decreasing the SFDR test result.

\begin{figure}[h]
\centering
\includegraphics[width=3.3in]{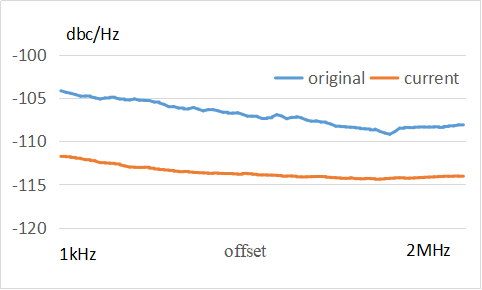}
\caption{Phase Noise Comparisioin of RF}
\label{RF}
\end{figure}

Test results for the phase noise of RF signal after the 250~MHz sine wave outputted from AWG mixing up with the 6~GHz signal generated from the microwave source are shown in Fig. \ref{RF}. 
The overall test results show a phase noise improvement of about 6~dB in the offset range of 2kHz to 2~MHz with respect to the original clock design. 
According to reference  \cite{Ball2016}, this result can improve the fidelity of qubits, which we will validate in future experiments. 

\section{Conclusion}
In this paper, a scalable self-adaptive synchronous triggering system is proposed to ensure synchronized operation of qubits in SQC. 
Firstly, an upgrated high-performacne clock system is designed to provide synchronized clocks for all AWGs, and it can bring a phase noise improvement of  about 6~dB for the control clock of qubits. 
The 250~MHz single-tone phase noise of DAC has been decreased about 15~dB.
Then a scalable master-slave star-like triggering method is proposed, which can realize a synchronization skew of more than 25~ps. 
And the synchronization of dozens or hundreds of qubits can be easily achieved. 
Finally, a self-adaptive solution which ensures the stable synchronization of multiple AWGs, is proposed to detect and calibrate the metastability. 
\bibliographystyle{IEEEtran}
\bibliography{Reference}
%

\end{document}